\documentstyle[12pt]{article}
\begin{document}
%
%
\newcounter{sec}
\def\be{\begin{eqnarray}}
\def\ee{\end{eqnarray}}
\def\ba{\begin{array}}
\def\ea{\end{array}}
\def\pa{\partial}
\def\ga{\gamma}
\def\sr{\stackrel{\rightarrow}}
\def\sl{\stackrel{\leftarrow}}
\def\srp{\stackrel{\rightarrow}{\partial}}
\def\slp{stackrel{\rightarrow}{\partial}}
\parskip=6pt	
\parindent=0pt
\renewcommand{\theequation}{\thesec.\arabic{equation}}
\baselineskip=22pt
\setcounter{sec}{2}
\bigskip\bigskip\bigskip\bigskip
\medskip
\begin{center}
\section*{ Wilson--Yukawa Chiral Model on Lattice
  and 
 Non-commutative 
Geometry  }

Jianming Li$^{a,b}$ , Xingchang Song$^{a,b}$ and Ke 
Wu$^b$
\\
$a$ Department of Physics, Peking University, Beijing 100871, 
China\footnote{Mailing address.
Email address: lijm@itp.ac.cn}\\and \\
$b$ Institute of Theoretical Physics, Academia Sinica, 
P.O.Box 2735, Beijing 100080, China
\end{center}



\begin{abstract}

 Based upon the mathematical  formulas of Lattice gauge theory and  
non-commutative geometry differential calculus, we developed  an approach of  
generalized gauge 
theory  on a product of the 
 spacetime lattice   and the two discrete points(or 
  a $Z_2$ 
discrete group).  We introduce a differentiation for non-nearest-neighbour 
points and 
find that this differentiation may lead to the introduction of Wilson term in 
the  free fermion Lagrangian  on lattice.
The Wilson-Yukawa chiral model on lattice is constructed by the generalized 
gauge 
theory and a toy model and  
Smit-Swift model are studied.
\end{abstract}

\vskip 1cm
{PACS number(s): 02.40. -k, 11.15. Ha}

\vfill
\newpage

\section[toc_entry]{ \bf Introduction}
Although there has been much activity in lattice gauge theory in past years, 
none of it 
has been concerned with its geometry framework until Dimakis 
et.al\cite{dimakis}  completed 
a non-commutative differential calculus on lattice in some sense of 
non-commutative differential geometry. Their results filled a space of 
understanding lattice gauge theory in geometry. In recent years, non-commutative 
geometry has brought a remarkable geometric picture explaining the nature of the 
Higgs field in the Standard Model\cite{connlot}---\cite{alvarez}. The 
electroweak part of the model is 
explained 
as originating from the product of the continuous Minkowski space-time by a 
discrete two-point space.  To study the Standard Model further, one must choose 
an appropriate regularization method. In fact, the lattice provides a very 
general 
regularization scheme for quantum field theories and as such has been applied to 
a variety of models. Several desirable models, one is so called Smit-Swift 
model\cite{smit,swift}, 
were proposed to stimulate the  Standard Model in lattice regularization 
technique. To study these models, fermion fields on lattice must be well 
defined. It is known that free fermion on lattice meet the ``species doubling'' 
problem. One  of the most popular schemes dealing with  the doubling problem was 
proposed by Wilson\cite{wilson}. In this approach, to get rid of the doubling 
problem, a new term which is called Wilson term was introduced technically in 
the free fermion lagrangian. This point was accepted extensively so far, but 
none can explain intrinsically why it exits. 

The purpose of this paper is to give a geometrical interpretation for the Wilson 
term and study how to construct the  
Wilson--Yukawa Chrial model on lattice from non-commutative differential 
geometry.  In section 2, we develop generalized differential calculus on the 
product space of spacetime lattice and  discrete two--point ( or a discrete 
group 
$Z_2$) and introduce a new kind of 
differentiation for 
non-nearest-neighbour points. Because non-commutative differential calculus of 
discrete two--point is 
equivalent to that of a discrete group $Z_2$ \cite{dimakis2}, we will don't 
distinguish them in 
the following discussions.   In section 3, first we will introduce a free 
fermion 
lagrangian on  the space of spacetime lattice product a discrete two-point.We 
find 
that the non-nearest-neighbour differentiations lead to the introduction of 
Wilson 
term. Then  we build generalized Yang-Mills-like 
gauge theory on spacetime 
lattice product discrete group $Z_2$ and construct a Wilson-Yukawa chiral model. 
Smit--Swift model as a special case of the 
model   is discussed in the section 4, then two examples, a toy model and 
electroweak model on lattice are studied.

\bigskip

\section[toc_entry]{ Differential calculus on  the product of space-time lattice 
and 
discrete group $Z_2$}
In this section, we develop the generalized   differential calculus on  the 
product of four 
dimensional  lattice 
and  two discrete points(or  discrete group $Z_2$) . For  details of 
differential calculus on lattice and discrete points, it refer to \cite{dimakis,
sitarz}.

We introduce a spacetime lattice with lattice spacing $a$ and two discrete 
points 
with spacing $1\over \mu$. Every point on the lattice is then specified by four 
integers and an element of discrete group $Z_2=\{e,Z| Z^2=e\}$ which we denote 
collectively by $(n,g)$, $n=(n_1,n_2,n_3,n_4), g\in Z_2$ and $n_i$ are integers.

Let ${\cal A}$ be the algebra of complex valued function on spacetime lattice 
and discrete group $Z_2$. An element of ${\cal A}$ may be written as 
$f(x,g),f\in {\cal A}$ , where $X=(x,g)$ is the coordinate , $x_\mu=n_\mu 
a(\mu=1,2,3,4)$. Actually, we may also understand discrete group $Z_2$ as an one 
dimensional periodic lattice with just two discrete  points.

The definition of  right action  operators on $\cal A$ is  as follows:
 $$R_\mu f(x,g)=f(x+a^\mu,g),\mu=1,2,3,4$$
 $$R_Z f(x,g)=f(x,g\cdot Z)$$
where  we used the notation
\be
(x+a^i )^j=x^j+\delta ^j_i a^i. \ee
It is obvious that 
$$R_\mu R_Z f(x,g)=R_Z R_\mu f(x,g)=f(x+a^\mu,g\cdot Z).$$

We introduce 
  discrete partial derivatives on   algebra $\cal A$ as 
follows:
 $$\ba{cl}
 &\pa_\mu f(x,g)=\frac 1 a (R_i f(x,g)-f(x,g))\\
 &\pa_Z f(x,g)=\mu (R_Z f(x,g)-f(x,g))\\
 &\pa_{Z\mu} f(x,g)=\frac 1 b (R_{\mu}R_Z f(x,g)-f(x,g)).\ea$$\\[5mm] 

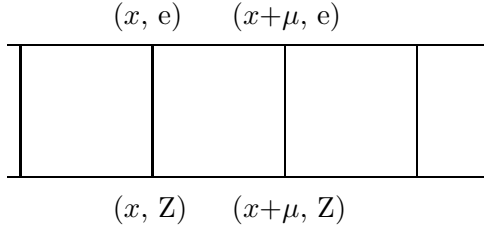
 \begin{picture}(50,50)(-50,0)
 \put(0,0){\line(1,0){180}}
 \put(0,50){\line(1,0){180}}
 \put(5,0){\line(0,1){50}}
 \put(55,0){\line(0,1){50}}
 \put(105,0){\line(0,1){50}}
 \put(155,0){\line(0,1){50}}
 \small{\put(40,-15){\makebox(0,2.5)[bl]{($x$, Z)}}
 \put(85,-15){\makebox(0,2.5)[bl]{($x$+$\mu$, Z)}}
 \put(40,58){\makebox(0,2.5)[bl]{($x$, e)}}
 \put(85,58){\makebox(0,2.5)[bl]{($x$+$\mu$, e)}}}
 \end{picture}\\[5mm]

 By the help of above picture, it is easy to show that we defined 
$\pa_\mu,\pa_Z$ by the partial difference 
between 
functions  of nearest neighbour points and $\pa_{Z\mu}$ by that of 
non-nearest-neighbour
points.In 
other words, we introduced  higher order derivatives on lattice,ie 
 $$
 b\pa_{Z\mu}={a\over \mu}\pa_\mu\pa_Z-a \pa_\mu-{1\over \mu }\pa_Z.$$
 Later we will find that the non-nearest-neighbour derivatives are useful to 
introduce Wilson fermion. In the Euclidean space, there is a relation among 
spacing  parameters  as
$$b^2=a^2+\frac 1 {\mu^2}.$$

In the following, we recall the notation of differential calculus on 
algebra $\cal A$, this is a Z-gradded algebra 
${\Omega}^{*}=\displaystyle\bigoplus_n{\Omega}^{n}$, ${\Omega}^{0}={\cal A}$
the elements of $\Omega^{n}$ are called n-forms.

To complete the construction of the differential calculus, we need to define the 
exterior derivative $d$, $d:~{\Omega}^{n} \rightarrow {\Omega}^{n+1}$
whose action on ${\cal A}$ is defined by
\be
df=\displaystyle{{\sum}_{\mu}}{\partial}_{\mu}f {dx}^{\mu}+\pa_Z f\chi+
\displaystyle{{\sum}_{\mu}}{\partial}_{Z\mu}f {\chi}^{\mu}, \ee
where $dx^\mu,\chi,\chi^\mu$ are basis of differential one form.
It is easy to show that  exterior derivative operator d 
 satisfies
$$\begin{array}{cl}
(i) ~~&d^{2}=0,\\[4mm]
(ii)~~&d(\omega_1 \omega_2)=d\omega_1\cdot 
\omega_2+(-1)^{deg\omega_1}\omega_1\cdot d\omega_2,~~~~~\forall \omega_1, 
\omega_2 \in {\Omega}^{*},
\end{array}$$
provided that $dx^\mu, \chi$ and ${\chi}^{\mu}$ satisfy the following  
conditions
\be
\begin{array}{cl}
dx^\mu f&=R_\mu f dx^\mu\\[4mm]
\chi f&=R_Z f \chi\\[4mm]
\chi^\mu f&= R_\mu R_Z f \chi^\mu\\[4mm]
d(dx^\mu)&=d\chi^\mu=0\\[4mm]
d\chi&=-2\chi\chi.
\end{array}\ee

As a direct results of the above formulas, we have that 
$$\ba{cl}
dx^\mu dx^\nu&=-dx^\nu dx^\mu\\[4mm]
\chi^\mu\chi^\nu&=-\chi^\nu\chi^\mu\\[4mm]
dx^\mu\chi&=-\chi dx^\mu\\[4mm]
dx^\mu\chi^\nu&=-\chi^\nu dx^\mu\\[4mm]
\chi^\mu\chi&=-\chi \chi^\mu.\ea$$

Let us now construct the generalized gauge theory on space-time lattice and 
discrete group 
using the above differential calculus.
We take the gauge transformations to be any proper subset ${\cal{H}}\subset
{\cal A}$. In particular, we will often take ${\cal H}$
to be unitary elements of ${\cal A}$,
\be
{\cal{H}}={\cal{U}}({\cal{A}})=\{a\in {\cal{A}}: 
~~~aa^{\dag}=a^{\dag}a=1\}.\ee
It is easy to see that the exterior derivative $d$ is not covariant with 
respect to 
the gauge transformations so that we should introduce the covariant 
derivative $d+A$, where $A$ is a generalized connection one-form. 
The requirement 
that $d+A$ is gauge covariant under gauge transformations
\be
 d+A \rightarrow H(d+A)H^{-1}, ~~~H\in \cal{H}, \ee
results in the following transformation rule of $A$, 
\be
A \rightarrow HA H^{-1}+HdH^{-1}. \ee 
If we write $A =\displaystyle{{\sum}_{\mu}}{A}_{\mu}{dx}^{\mu}+\phi\chi+
\displaystyle{{\sum}_{\mu}}{B}_{\mu}{\chi}^{\mu}$, 
under gauge transformation $H\in \cal{H}$, 
${A}_{\mu},\phi,B_\mu$ transform as
\be\ba{cl}
 {A}_{\mu}& \rightarrow 
H{A}_{\mu}({R}_{\mu}H^{-1})+H{\partial}_{\mu}H^{-1}\\[4mm]
 \phi& \rightarrow H{\phi}_{\mu}({R}_{\mu}H^{-1})+H{\partial}_{Z}H^{-1}\\[4mm]
 {B}_{\mu}& \rightarrow H{B}_{\mu}({R}_{\mu}R_Z 
H^{-1})+H{\partial}_{Z\mu}H^{-1}.\ea
\label {A}\ee
It is convenient to introduce a new field varibles  
$$\ba{cl}
G_\mu&=1+aA_\mu, \mu=1,2,3,4\\[4mm]
\Phi&=\mu+ 
 \phi\\[4mm]
K_\mu&=1+bB_\mu, \mu=1,2,3,4. 
\ea$$
 Then 
(\ref A) is equivalent to 
\be\ba{cl}
{G}_{\mu} &\rightarrow H{G}_{\mu}({R}_{\mu}H^{-1})\\[4mm]
\Phi &\rightarrow H\Phi R_Z H^{-1}\\[4mm]
{K}_{\mu} &\rightarrow H{K}_{\mu}({R}_{\mu} R_Z H^{-1}).\ea\label{gauge} \ee

It can be shown that the generalized curvature two form
\be
 F=d\phi+\phi \otimes \phi \ee
is gauge covariant and can be written in terms of its  coefficients 
\be\ba{cl}
 F=&\displaystyle{\sum_{\mu,\nu}}F_{\mu,\nu}{dx}^{\mu}\otimes{dx}^{\nu} 
+\displaystyle{\sum_{\mu,\nu}}F_{Z\mu,Z\nu}{\chi}^{\mu}\otimes{\chi}^{\nu}\\[4mm
]
 
&+\displaystyle{\sum_{\mu,\nu}}F_{\mu, Z\mu}{dx}^{\mu}\otimes{\chi}^{\nu}+
\displaystyle{\sum_{\mu}}F_{\mu, Z}{dx}^{\mu}\otimes{\chi}\\[4mm]
&+\displaystyle{\sum_{\mu}}F_{Z\mu, Z}{\chi}^\mu\otimes{\chi}+
F_{Z, Z}\chi\otimes{\chi},\ea\ee
where 
\be\ba{cl}
 F_{\mu,\nu}&=\frac 1 2 \frac 1 {a^2} ({G}_{\mu}R_\mu 
{G}_{\nu}-{G}_{\nu}{R}_{\nu}{G}_{\mu})\\[4mm]
 F_{Z\mu,Z\nu}&=\frac 1 2 \frac 1 {b^2} ({K}_{\mu}R_\mu 
R_Z{K}_{\nu}-{K}_{\nu}{R}_{\nu}R_Z{K}_{\mu})\\[4mm]
  F_{\mu,Z\mu}&= \frac 1 {ab} ({G}_{\mu}R_\mu 
K_\nu-K_\nu R_\mu R_Z{G}_{\mu})\\[4mm]
F_{\mu,Z}&=\frac 1 a (G_\mu R_\mu \Phi-\Phi R_Z G_\mu)\\[4mm]
F_{Z\mu,Z}&=\frac 1 b (K_\mu R_\mu \Phi-\Phi R_Z K_\mu)\\[4mm]
F_{Z,Z}&=(\Phi R_Z \Phi-\mu^2).\ea\ee

In order to get Lagrangian of the Yang-Mills type, we need to define a metrices 
as 
follows,
\be\ba{cl}
&<{dx}^{\mu},{dx}^{\nu}>=g^{\mu\nu},~~~~<{\chi},{\chi}>=
{\eta},~~~~<\chi^{\mu},\chi^\nu>=\xi g^{\mu,\nu}\\[4mm]
&<dx^\mu,\chi^\nu>=<\chi^\mu,dx^\nu>=0,\\[4mm]
&<dx^\mu,\chi>=<\chi,dx^\nu>=0,~~~
<\chi^\mu,\chi>=<\chi,dx^\nu>=0\\[4mm]
&<{dx}^{\mu}\wedge {dx}^{\nu},{dx}^{\sigma}\wedge{dx}^{\rho}>=
\frac{1}{2}(g^{\mu\sigma}g^{\nu\rho}-g^{\mu\rho}g^{\nu\sigma}),\\[4mm]
&<{\chi}^{\mu}\wedge {\chi}^{\nu},{\chi}^{\sigma}\wedge{\chi}^{\rho}>=
\xi^2\frac{1}{2}(g^{\mu\sigma}g^{\nu\rho}-g^{\mu\rho}g^{\nu\sigma}),\\[4mm]
&<{dx}^{\mu}\otimes{\chi}^{p},{dx}^{\nu}\otimes {\chi}^{q}>=
\xi g^{\mu\nu} g^{pq},~~~<{dx}^{\mu}\otimes{\chi},{dx}^{\nu}\otimes {\chi}>=
g^{\mu\nu} \eta\\[4mm]
& <{\chi}^{\mu}\otimes{\chi},{dx}^{\nu}\otimes {\chi}>=
\xi g^{\mu\nu} \eta,~~~<{\chi}\otimes {\chi},{\chi}\otimes {\chi}>=
{\eta}^2,\ea \ee
 Thus, we may introduce the Lagrangian of the Yang-Mills type 
for the gauge boson and Higgs from the inner product of curvature 
$<F,\overline{F}>$, where 
$$\ba{cl}
\overline{F}=
&\displaystyle{\sum_{\mu,\nu}}{dx}^{\mu}\otimes{dx}^{\nu}F^{\dag}_{\mu,\nu} 
+\displaystyle{\sum_{\mu,\nu}}{\chi}^{\mu}\otimes{\chi}^{\nu}F^{\dag}_{Z\mu,Z\nu
}\\[4mm
]
 &+\displaystyle{\sum_{\mu,\nu}}{dx}^{\mu}\otimes{\chi}^{\nu}F^{\dag}_{\mu, 
Z\mu}+
\displaystyle{\sum_{\mu}}{dx}^{\mu}\otimes{\chi}F^{\dag}_{\mu, Z}\\[4mm]
&+\displaystyle{\sum_{\mu}}{\chi}^{\mu}\otimes{\chi}F^{\dag}_{Z\mu, Z}+
\chi\otimes{\chi}F^{\dag}_{Z, Z}.\ea$$

After cumbersome calculation, we get the Lagrangian as follows:
\be\ba{cl}
&{\cal L}_{YM-H}
=-\frac 1 N <F,\overline{F}>\\[4mm]
=&
{1\over a^4}\displaystyle\sum_{\mu<\nu}[ 1-\frac 1 2 
(U_{\mu,\nu}+U^{\dag}_{\mu,\nu})]\\[4mm]
&+{\xi^2\over  b^4}\displaystyle\sum_{\mu\nu}\frac 1 2 [K^{\dag}_\mu 
K_\mu R_\mu R_Z (K_\nu K^{\dag}_\nu)-\frac 1 2 ([K^{\dag}_\mu 
K_\nu R_Z(R_\nu K_\mu R_\mu K^{\dag}_\nu)+K\nu^{\dag} K_{\mu}R_Z(R_\mu K_\nu 
R_\nu K^{\dag}_\mu)]\\[4mm]
&{\xi \over a^2b^2 }\displaystyle\sum_{\mu\nu} [
2K_\mu K^{\dag}_\mu-(G_\mu R_\mu K_\nu R_\nu R_Z G_\mu^{\dag} 
K^{\dag}_\nu+K_\nu R_\nu R_Z G_\mu R_\mu K^{\dag}_\nu G^{\dag}_\mu)]\\[4mm]
&+{\eta\over a^2}\displaystyle\sum_{\mu} [
2\Phi \Phi^{\dag} -(G_\mu R_\mu \Phi R_Z G^{\dag}_\mu \Phi^{\dag}+
\Phi R_Z G_\mu R_\mu \Phi^{\dag} G^{\dag}_\mu)]\\[4mm]
&+{\xi \eta  \over b^2}\displaystyle\sum_{\mu} [
K^{\dag}K_\mu R_\mu R_Z (\Phi\Phi^{\dag})+\Phi^{\dag}\Phi R_Z K_\mu K^{\dag}-
K_\mu R_\mu R_Z \Phi R_Z K^{\dag}_\mu \Phi^{\dag}-\Phi R_Z K_\mu R_\mu R_Z 
\Phi^{\dag} K^{\dag}_\mu]\\[4mm]
&+\eta^2  [\Phi \Phi^{\dag} -\mu^2]^2\ea\ee
where $~~~U_{\mu,\nu}=G_\mu R_\mu G_\nu R_\nu 
G_\mu^{\dag} G_\nu^{\dag}$.

Hence, we have got the lagrangian of gauge fields on each site, which is a 
function of coordinate $(n,g)$. To obtain the physical action of gauge fields, 
we need to integrate ${\cal L}$ over spacetimes lattice and discrete group $Z_2$ 
$$S_G=a^4 \displaystyle\sum_{n,g}{\cal L}.$$
The exact expression of $S_G$ will be given in next section. There are three 
types of gauge field in the generalized gauge theory, two of 
them---$G_\mu,K_\mu$ 
should be vectorlike and $\Phi$ should be scalar. As a special case, it is 
possible to express $K_\mu$ in terms of $G_\mu$ and $\Phi$, this will be 
discussed in construction of physical model in next section.

\setcounter{equation}{0}
\setcounter{sec}{3}
 \section[toc_entry]{ Generalized Gauge Theory on Lattice } 
  We have constructed a generalized differential calculas  on lattice and 
shown that the Higgs fields and the vectorlike fields may be introduced as 
generalized 
gauge fields on spacetime and 
discrete group as well as Yang-Mills fields on 4-dimmensional space time 
lattice. To deal with the  model building in particle physics, fermions and 
their couplings to gauge fields must be included , we will first complete this 
issues in this section and then build the generalized gauge theory.

We now consider fermion fields on space time lattice and discrete group $Z_2$. 
In space time lattice theory, it is known that the fermion species doubling 
problem must be suppressed with some appropriate modification of the 
latticedized theory. One of the successful approaches as proposed by Wilson is 
so 
called Wilson fermion. In this 
paper, we first introduce Wilson-like fermion on space time lattice 
and discrete 
group $Z_2$
 by taking into account the non-nearest-neighboure derivative of last section. 
 Frist we introduce  the free fermion lagrangian on lattice as follows
\be\ba{cl}
{\cal L}(n,g)=&\overline \psi(n,g)\ga^\mu(\stackrel{\rightarrow}{\pa_\mu } -
\sl{\pa_\mu})\psi(n,g)- \overline\psi(n,g)\pa_Z \psi(n,g) \\[4mm]
&+ \frac q a  \overline 
\psi(n,g)\displaystyle\sum_{\mu}(\stackrel{\rightarrow}{\pa}_{Z\mu 
}  +\sl{\pa}_{Z\mu} )\psi(n,g),\ea ~~~g\in Z_2 
\label{fl1}\ee
where $\psi(e)=\psi_L$, $\psi(Z)=\psi_R$ are left  and right handed ferimons 
respectively. It is noted that the first and second terms in the lagrangian are 
similar as those  of  previous works\cite{mine1,mine2}, 
if we want to include the 
the non-neighbouring differentiation $\pa_{Z\mu}$ in the lagrangian , the third 
term is in its  most simple and non-trivial forms. The factor $\frac 1 a$ is 
need to cancall the doubling. 

It is easy to show that we can get the action of Wilson fermion when integrating 
the lagrangian (\ref {fl1}), 
\be\ba{clll}
S_F&=&a^4\displaystyle\sum_{n,g}{\cal L}(n,g)\\[4mm]
&=&a^3 \displaystyle\sum_{n,\mu}[\overline\psi (n)
\ga^\mu \psi(n+\mu)-\overline\psi (n+\mu)
\ga^\mu \psi(n)]+a^4\displaystyle\sum_{n}\mu\overline\psi(n) 
\psi(n)\\[4mm]
&&+ \frac {q} b a^3 
\displaystyle\sum_{n,\mu}[\overline\psi(n)\psi(n+\mu)+\overline\psi(n+\mu)
\psi(n)-2\overline\psi(n)\psi(n)].\ea 
\ee

So in the following, we will build lattice gauge theory using the lagrangian 
(\ref 
{fl1}).
The lagrangian  (\ref {fl1}) is invariant under the global transformations
\be\ba{cl} 
&\psi(n)\rightarrow G \psi(n),\\
&\overline\psi (n)\rightarrow \overline \psi(n) G^{-1}\ea \label{gauge1}\ee
where G is an element of the gauge group.
Similar to the reason that leads to the introduction of Yang-Mills fields,
it is reasonable to require that the Lagrangian (\ref {fl1}) be invariant under 
local gauge 
transformations $H(n,h), ~h\in Z_2$,
\be
\psi(x,h)\rightarrow {\psi(x,h)}'=H(x,h)\psi(x,h). \ee
Then the discrete partial  dervative in (\ref {fl1}) should be replace by 
covariant derivative as follows 
\be
\pa_Z \rightarrow D_Z,~~~\sr{\pa}_i\rightarrow \sr{D_i}, 
~~~\sl{\pa_i}\rightarrow \sl{D_i}, ~~~i=\mu, Z\mu,\ee
and $\sr{D}_i,\sl D_i$ and $D_Z$ have the simple transformations under gauge 
transformation (\ref {gauge1}),
\be\ba{cl}
&\sr {D_i} \psi(n,g)\rightarrow[\sr {D_i} \psi(n,g)]^\prime=H(n,g) \sr 
{D_i}\psi(n,g)\\[4mm]
& \overline\psi(n,g) \sl {D_i}\rightarrow[\overline\psi(n,g) \sl 
{D_i}]^\prime=\overline\psi(n,g) \sl {D_i}H^{-1}(n,g) \\[4mm]
&{D_Z} \psi(n,g)\rightarrow[{D_Z} \psi(n,g)]^\prime=H(n,g)  
{D_Z}\psi(n,g),\ea\ee
where the covariant derivatives are formed as 
\be\ba{cl}
\sr{D_\mu} \psi(n,g)&=( {\pa_\mu}+ig A_\mu {R_\mu} )\psi(n,g)={1\over a}(U_\mu  
{R_\mu}-1)\psi(n,g)\\[4mm]
\sr{D_{Z\mu}} \psi(n,g)&=( {\pa_{Z\mu}}+ig B_\mu {R_\mu} R_Z)\psi(n,g)={1\over 
b} (K_\mu 
{R_\mu} R_Z-1)\psi(n,g)\\[4mm]
{D_Z} \psi(n,g)&=( {\pa_Z}+\lambda \phi R_Z )\psi(n,g)=\lambda(\Phi 
R_Z-{\mu\over \lambda})\psi(n,g),\ea\ee
where
$$\ba{cl}
G_\mu&=1+ig aA_\mu, \mu=1,2,3,4\\[4mm]
\Phi&={\mu\over \lambda}+ 
 \phi\\[4mm]
K_\mu&=1+bB_\mu, \mu=1,2,3,4 
\ea$$
correspondingly, we have the expression of $\sl{D_i}$ as
$$\ba{cl}
\overline\psi(n,g) \sl{D_\mu}&=\frac 1 a \overline \psi(n,g)(\sl{R_\mu 
}U^{\dag}_\mu-1)\\[4mm]
\overline\psi(n,g) \sl{D}_{Z\mu}&=\frac 1 b \overline \psi(n,g)(\sl{R_{Z\mu 
}}K^{-1}_\mu-1).\ea$$
Then the transformation law (\ref {gauge1}) are satisfied if the generalized 
gauge fields $U_\mu,K_\mu$ and $\Phi$ have the following properties:
\be\ba{cl}
{G}_{\mu} &\rightarrow H{G}_{\mu}({R}_{\mu}H^{-1})\\[4mm]
\Phi &\rightarrow H\Phi R_Z H^{-1}\\[4mm]
{K}_{\mu} &\rightarrow H{K}_{\mu}({R}_{\mu} R_Z H^{-1}).\ea\label{gauge2}\ee

Similar to the usual gauge theory, the covariant dreivative is equivalent to the 
covariant exterior derivative $D=d+A$ and  
\be
Df=D_\mu f dx^\mu +D_{Z\mu} \chi^\mu +D_Z f \chi.\ee

Thus from (\ref {fl1}), the gereralized gauge invariant lagrangian for fermions 
should be,
\be\ba{cl}
{\cal L}_F (n,g)=&\overline \psi(n,g)\ga^\mu (\stackrel{\rightarrow}{D_\mu } -
\sl{D_\mu})\psi(n,g)- \overline\psi(n,g)D_Z \psi(n,g)\\[4mm]
& +  \frac q a \overline 
\psi(n,g)\displaystyle\sum_{\mu}(\stackrel{\rightarrow}{D}_{Z\mu 
}  +\sl{D}_{Z\mu} )\psi(n,g),\ea 
\label{flc}\ee
and in terms of the expessions of $D_i, D_Z$, it may be written as:
\be\ba{cl}
{\cal L}(n,g)=&\frac 1 a \displaystyle\sum_{\mu} [\overline \psi(n,g)\ga^\mu 
{G_\mu(n,g) }\psi(n+\mu,g) 
-
\overline \psi(n+\mu,g)\ga^\mu G_\mu^{\dag} \psi(n,g)]\\[4mm]
&+\lambda 
\overline\psi(n,g)\Phi(n,g) 
\psi(n,gZ) \\[4mm]
&+ \frac q {ab} \displaystyle\sum_{\mu}[\overline 
\psi(n,g) K_\mu\psi(n+\mu, gZ)+\overline\psi(n+\mu,gZ) K^{\dag}_\mu\psi(n, 
gZ)].\ea
 \label{flc1}\ee

In the following calculaiton, we set 
\be\ba{cl}
G_\mu(n,e)=G^L_\mu,&G_\mu(n,Z)=G^R_\mu(n)\\[4mm]
\Phi(n,e)=\Phi(n),&\Phi(n,Z)=\Phi(n)\\[4mm]
K_\mu(n,e)=K^1_\mu(n),&K_\mu(n,Z)=K^2_\mu(n).\ea\ee
After integrating lagrangian (\ref {flc1}) over spacetime and discrete group 
$Z_2$, we get the covariant action of fermoin sector,
\be\ba{cl}
S_F=&a^3\displaystyle\sum_{n,\mu}[\overline \psi_L(n)\ga^\mu 
{G^L_\mu(n) }\psi_L(n+\mu,g) -
\overline \psi_L(n+\mu,g)\ga^\mu G^{L\dag}_\mu \psi_L(n,g)\\[5mm]
&~~~~~~+\overline \psi_R(n)\ga^\mu {G^R_\mu(n) }\psi_R(n+\mu,g) -
\overline\psi_R(n+\mu,g)\ga^\mu G^{\dag}R_\mu \psi_R(n,g)]\\[5mm]
&+ \frac {q} b  a^3\displaystyle\sum_{n,\mu}[\overline 
\psi_L(n) K^1_\mu\psi_R(n+\mu)+\overline\psi_R(n) 
K^{2}_\mu\psi_L(n+\mu)+h.c.]]\\[5mm]
&+\lambda a^4\displaystyle\sum_{n}[\overline\psi_L(n)\Phi(n) \psi_R(n) 
+\overline\psi_R(n)\Phi^{\dag}(n) \psi_L(n)
].\ea
\label{fl3}\ee

Using the results of geralized differential calculus on space time lattice and 
discrete group $Z_2$ of last section, we may introduce the action of gauge 
fields from the curvature as follows:
\be\ba{cl}
S_G=&\displaystyle\sum_{n,\mu,\nu}\frac 1 2 
[Tr(1-U^L_{\mu,\nu})+Tr(1-U^R_{\mu,\nu})]\\[4mm]
&+{\xi^2\over b^4} a^4\displaystyle\sum_{n,\mu,\nu}\frac 1 2
Tr[K^{1\dag}_\mu K^1_\mu R_\mu(K^2_{\nu}K^{2\dag}_\nu)-
K^{1\dag}_\mu K^1_\nu R_\nu K^2_{\mu} R_\mu K^{2\dag}_\nu]\\[4mm]
&+{\xi \over a^2 b^2}a^4\displaystyle\sum_{n,\mu,\nu}
Tr[2(K^{1}_\mu K^{1\dag}_\mu +K^2_{\mu}K^{2\dag}_\mu)-
(G^L_\mu R_\mu K^{1}_\nu  R_\nu G^{R\dag}_\mu K^{1\dag}_{\mu}+
G^R_\mu R_\mu K^{2}_\nu  R_\nu G^{L\dag}_\mu K^{2\dag}_{\mu}+h.c.) ]\\[4mm]
&+{\xi \eta\over  a^2}a^4\displaystyle\sum_{n,\mu}
[4\Phi\Phi^{\dag}-2(G^L_\mu R_\mu\Phi G^{R\dag} \Phi^{\dag}+h.c.)]\\[4mm]
&+{\xi \eta  \over b^2} a^4\displaystyle\sum_{n,\mu} [
K^{1\dag}_\mu K^1_\mu R_\mu (\Phi^{\dag}\Phi)+\Phi^{\dag}\Phi K^2_\mu 
K^{2\dag}_\mu+
K^{2\dag}_\mu K^2_\mu R_\mu (\Phi\Phi^{\dag})+\Phi\Phi^{\dag} K^1_\mu 
K^{1\dag}_\mu\\[4mm]
&~~~~~~-
2K^1_\mu R_\mu\Phi^{\dag} K^{2\dag}_\mu \Phi^{\dag}-\Phi K^2_\mu R_\mu
\Phi K^{1\dag}_\mu]\\[4mm]
&+\eta^2 a^4\displaystyle\sum_{n} 2 Tr(\Phi\Phi^{\dag}-{\mu^2\over 
\lambda^2})^2.\ea\ee

So far, we have  completed the constructin of generalized gauge theory on 
spacetime 
lattice and discrete group $Z_2$. In next section, we will show the Smit-Swift 
model on lattice is a special case of  the model we dicussed above.

\setcounter{sec}{4}\setcounter{equation}{0}
\section[toc_entry]{ Smit-Swift Model } 

We have finished the construction of physical model in last subsection, three 
kinds of gauge field---$G_\mu,K_\mu, \Phi$ are introduce in the approach, but, 
in realistic lattice 
model, there are only two kinds of gauge fields ---Yang-Mills fields $G_\mu$ and 
Higgs fields $\Phi$, so 
it is desired to express the third gauge fields $K_\mu$ in terms of $G_\mu$ and 
$\Phi$ or we treat the third connection as a combination of the others.
 By the requirement of gauge transformation 
properties (\ref {gauge2}), the most general experssion of $H_\mu$ in terms of 
$G_\mu$ and $\Phi$ is the linear combination of two terms $G_\mu R_\mu \Phi$
and 
$\Phi R_Z G_\mu$, and the most general expression is 
\be\ba{cl} 
&K^1_\mu=\alpha_1 \Phi G^R_\mu+\beta_1 G^L_\mu R_\mu \Phi\\[4mm]
&K^2_\mu=\alpha_2 \Phi^{\dag} G^L_\mu+\beta_2 G^R_\mu R_\mu \Phi^{\dag}.\ea
\label{exp1}\ee

In the following, we will find that Smit-Swift model may be reconstructed when   
a special case of (\ref {exp1}) is taken into account,
\be\ba{cl} 
&K^1_\mu=f\Phi G^R_\mu
,~~~K^2_\mu=f G^R_\mu R_\mu \Phi^{\dag},\ea\ee where $f$ is a free parameter.

Substitute the expression of $K^1_\mu$ and $K^2_\mu$ into the fermion 
action (\ref {fl3}), one obtains
\be\ba{cl}
S_F=&a^3\displaystyle\sum_{n,\mu}[\overline \psi_L(n)\ga^\mu 
{G^L_\mu(n) }\psi_L(n+\mu,g) -
\overline \psi_L(n+\mu,g)\ga^\mu G^{L\dag}_\mu \psi_L(n,g)\\[5mm]
&~~~~~~+\overline \psi_R(n)\ga^\mu {G^R_\mu(n) }\psi_R(n+\mu,g) -
\overline\psi_R(n+\mu,g)\ga^\mu G^{\dag}R_\mu \psi_R(n,g)]\\[5mm]
&+ r a^3 \displaystyle\sum_{n,\mu}[\overline 
\psi_L(n) \Phi G^R_\mu\psi_R(n+\mu)+\overline\psi_R(n)G^R_\mu \Phi(n+\mu) 
\psi_L(n+\mu)+h.c.]]\\[5mm] 
&+\lambda a^4 \displaystyle\sum_{n}[\overline\psi_L(n)\Phi(n) \psi_R(n) 
+\overline\psi_R(n)\Phi^{\dag}(n) \psi_L(n)
],\ea\label{fermion}\ee where $r=\frac q b f$.

It is easy to show that the Higgs potential reach its minimu when $|\Phi|=\frac 
\mu \lambda$. To simplify the problem( without throwing away any important 
physics \cite{frad, swift}), we freeze 
out the radial model of the Higgs fields, working with  fields with fixed norm 
$\Phi\Phi^{\dag}=\frac {\mu^2} {\lambda^2}$. Thus we shall be dealing with 
fields that are strictly 
compact. 
Under above 
assumption, we obtain the Lagrangian of gauge field by throwing away constant 
numbers as following:

\be\ba{cl}
S_G=&\frac 1 2  Tr \displaystyle\sum_{n,\mu,\nu}
 U^L_{\mu,\nu}+\frac 1 2 (1 +{\xi^2 f^4} {\mu^4\over \lambda^4}{a^4 \over 
b^4})Tr 
\displaystyle\sum_{n,\mu,\nu}
U^R_{\mu,\nu}\\[4mm]
&+{f^2 \over  b^2} a^2\xi \displaystyle\sum_{n,\mu,\nu}
Tr[U_{\mu,\nu} S_\mu+{\tilde U}^R_{\mu,\nu} R_\nu 
S^{\dag}_\mu)+h.c.]
 ]\\[4mm]
&+2{\xi \eta a^2} \displaystyle\sum_{n,\mu}
 (S_\mu+S^{\dag}_\mu)],\ea\ee
where 
$S_\mu=\Phi^{\dag} G_\mu(e) R_\mu \Phi G^{\dag}_\mu(Z)
$, ${\tilde U}^R_{\mu,\nu}=G^{R\dag}_\nu G_\mu R_\mu G_\nu R_\nu 
G^{R\dag}_\mu$. It is easy to show that ${\tilde U}^R_{\mu,\nu}={ 
U}^R_{\mu,\nu}$ if the gauge fields $U^R$ is Abelian. Parameters $f,\xi, \eta$ 
are free, so up to a proportionate constant, we may write the lagrangian in the 
following form:
\be\ba{cl}
S_G=&\beta_1 Tr \displaystyle\sum_{n,\mu,\nu}
 U^L_{\mu,\nu}+\beta_2 Tr 
\displaystyle\sum_{n,\mu,\nu}
U^R_{\mu,\nu}\\[4mm]
&+\beta_3 a^2\displaystyle\sum_{n,\mu,\nu}
Tr[U_{\mu,\nu} S_\mu+{\tilde U}^R_{\mu,\nu} R_\nu 
S^{\dag}_\mu+h.c.
 ]\\[4mm]
&+\beta_H a^2\displaystyle\sum_{n,\mu}
 (S_\mu+S^{\dag}_\mu)],\ea\label{bosonic}\ee 
where 
$\beta_1,\beta_3, \beta_H$ are free parameters and $\beta_2={\beta^2_1+\beta^2_3 
 a^4\over \beta_1}$.

There is a high order 
terms appear in the lagrangian, which is different from ordinary lattice gauge 
theory, 
whether this term can lead to new physics need to be studied future.

\subsection[toc_entry]{ A Toy Model } 

Now we consider a ``toy model'' in which left-handed fields transform according 
to the fundamental representation of a gauge group---say $SU(2)$---and there is 
a right-handed partner which transforms trivially. Hence, the gauge fields  
$G_\mu$ and $\Phi$ on each sector should be: 
\be\ba{cl}
G^L_\mu=U_\mu,&G^R_\mu=1\\[4mm]
K^1_\mu=f\Phi ,&K^2_\mu=f R_\mu \Phi.\ea\ee

As a direct results of (\ref {fermion})
\be\ba{cl}
S_F=&a^4 \displaystyle\sum_{n,\mu}[\overline \psi_L(n)\ga^\mu {U_\mu(n) 
}\psi_L(n+\mu,g) 
-
\overline \psi_L(n+\mu,g)\ga^\mu U^{\dag}_\mu \psi_L(n,g)\\[5mm]
&~~~~~~+\overline \psi_R(n)\ga^\mu \psi_R(n+\mu,g) -
\overline \psi_R(n+\mu,g)\ga^\mu  \psi_R(n,g)]\\[5mm]
&+\lambda a^4\displaystyle\sum_{n}[\overline\psi_L(n)\Phi(n) \psi_R(n) 
+\overline\psi_R(n)\Phi^{\dag}(n) 
\psi_L(n)] \\[5mm]
&+ ra^3 \displaystyle\sum_{n,\mu}[\overline 
\psi_L(n) \Phi\psi_R(n+\mu)+\overline\psi_R(n)  
\Phi^{\dag}(n+\mu)\psi_L(n+\mu)+h.c.]\ea\ee

Now  freezing   out the radial model of the Higgs fields and throwing away some 
constant, we get a simple mode
\be
S_G=\beta \displaystyle\sum_{n,\mu,\nu}
[Tr U_{\mu,\nu})]+
\beta_H Tr[\Phi^{\dag}U_\mu R_\mu\Phi  +h.c.].\ee
where $\beta$ and $\beta_H$ are free parameters.
This toy model is study in many works, such as \cite{swift}.

\subsection[toc_entry]{$SU(2)\times U(1)$ Electroweak Model on Lattice } 

With all previous points in mind, a construction  for the $SU(2)\times U(1)$ 
electroweak theory on the lattice is now available. Consider just the leptonic 
first generation sector: electron and electron neutrino and define the fields as 
follows:
 
\be\ba{clll}
&\psi(e)=L=\left( \ba{cl}v_L\\e_L\ea\right),~~~~~~ &\psi(Z)=\left( 
\ba{cl}R_2\\R_1\ea\right)=\left( \ba{cl}v_R\\e_R\ea\right) \\[10mm]
&G_\mu(e)=G_\mu^L=U_\mu V_\mu, 
&G_\mu(Z)=G_\mu^R=\left(\ba{cc}1&\\&V'_\mu\ea\right)\\[10mm]
&\Phi(e)=\left(\ba{cc}\phi_0^*&\phi_+\\-\phi_+^*& \phi_0\ea\right),
&\Phi(Z)=\left(\ba{cc}\phi_0&-\phi_+\\\phi_+^*& \phi_0^*\ea\right)\\[10mm]
&K^1_\mu=\Phi G_\mu(Z) f,&K^2_\mu=fG_\mu(Z) R_\mu \Phi^{\dag}  \ea 
\label{asign}\ee
where $U_\mu(n)=exp[iga\tau\cdot A_\mu(n)]$ is $SU(2)$ gauge field and  
$V_\mu(n)=exp[- {1\over 2} 
ig'aB_\mu(n)]$, $V^\prime_\mu(n)=exp[-  
ig'aB_\mu(n)]$ are $U(1)$ gauge field, $f=\left(\ba{cc}f_2&\\&f_1\ea\right)$ is 
introduced as non-trivial coupling constant which is consistent with gauge 
transformation properties (\ref {gauge2}). 

Using the results of last section and those assigments of (\ref{asign}), we have 
the action for the fermion part,
\be\ba{cl}
L=&a^3 \displaystyle\sum_{n}[\overline L(n)\gamma^\mu U_\mu V_\mu 
L(n+\mu)-\overline 
L(n+\mu)\gamma^\mu U^{\dag}_\mu 
V^{\dag}_\mu L(n)\\
&~~~~~~+\overline R_1(n)\gamma^\mu  V'_\mu R_1(n+\mu)-\overline 
R_1(n+\mu)\gamma^\mu 
V'^{\dag}_\mu 
R_1(n)\\
&~~~~~~+\overline R_2(n)\gamma^\mu  R_1(n+\mu)-\overline R_1(n+\mu)\gamma^\mu 
R_1(n)]\\
&+ a^4 \displaystyle\sum_{n}\lambda [\overline L(n)\phi R_1+\overline 
R_1(n)\phi^{\dag} L(n)+\overline 
L(n)\stackrel{\sim}{\phi} R_2+\overline R_2(n)\stackrel{\sim}{\phi}^{\dag} 
L(n)]\\
&+a^3\displaystyle\sum_{n,\mu}r_1 [\overline L(n)V'_\mu(n)\phi 
R_1(n+\mu)+\overline 
R_1(n)V'_\mu(n)\phi^{\dag}(n+\mu) L(n+\mu)+h.c.]\\
& a^3\displaystyle\sum_{n,\mu}r_2 [\overline L(n)\stackrel{\sim}{\phi} 
R_2(n+\mu)+\overline 
R_2(n)\stackrel{\sim}{\phi}^{\dag}(n+\mu) L(n+\mu)+h.c.]\ea\ee
where $\phi=\left(\ba{cl}\phi_+\\\phi_0\ea\right)$ and 
$\tilde{\phi}=\left(\ba{cl}\phi^*_0\\ -\phi^*_+\ea\right)$, two constants 
are $r_1=f_1 \frac {q} b$, $r_2=f_2 \frac {q} b$.

Substitute the assignments of fields (\ref {asign}) into action of bosonic 
sector(\ref {bosonic}), we have 
\be\ba{cl}
S_G=
& \beta_1 Tr \displaystyle\sum_{n,\mu,\nu}
 U_{\mu,\nu}+\beta_2 Tr 
\displaystyle\sum_{n,\mu,\nu}
V^{\prime}_{\mu,\nu}+\beta_H \displaystyle\sum_{n,\mu} Tr S_\mu\\[4mm]
&+\beta_3Tr \displaystyle\sum_{n,\mu,\nu}
\left[\left(\ba{cc}({r_2\over r_1})^2&\\&{ 
V}^{\prime}_{\mu,\nu}\ea\right)(S_\mu+R_\nu 
S^{\dag}_\mu)+\left(\ba{cc}({r_2\over r_1})^2&\\&{ 
V}^{\prime\dag}_{\mu,\nu}\ea\right) (S^{\dag}_\mu+R_\nu 
S_\mu)\right],\ea\ee
where
$$S_\mu=\Phi^{\dag} U_\mu 
R_\mu\Phi  \left(\ba{cc}  V^{\dag}_\mu&\\& V_\mu\ea\right).$$
It is easy to show that 
\be
Tr[S_\mu]
=\phi^{\dag}(n)U_\mu(n)V_\mu^{\dag}\phi(n+\mu)+\phi^{\dag}(n+\mu)U^{\dag}_\mu 
V^{\dag}_\mu\phi(n),\ee
which is the kinetic term of Higgs field on lattice in Smit-Swift 
model\cite{swift}.

At last  we can write the lagrangian of bosonic part as following:
 \be\ba{cl}
S_G=
& \beta_1 Tr \displaystyle\sum_{n,\mu,\nu}
 U_{\mu,\nu} V_{\mu,\nu}+\beta_2 Tr 
\displaystyle\sum_{n,\mu,\nu}
V^{\prime}_{\mu,\nu}\\[4mm]
&+\beta_H a^2 \displaystyle\sum_{n,\mu} 
[\phi^{\dag}(n)U_\mu(n)V_\mu^{\dag}(n)\phi(n+\mu)+
\phi^{\dag}(n+\mu)U^{\dag}_\mu(n) 
V_\mu(n)\phi(n)]\\[4mm]
&+\beta_3 a^2 Tr \displaystyle\sum_{n,\mu,\nu}\left[\phi^{\dag}(n) 
U_\mu(n) 
V^{\dag}_\mu(n) \Phi(n+\mu) [V^{\prime}_{\mu,\nu}(n)+ 
V^{\prime\dag}_{\mu,\nu}(n-\nu)]
+ h.c.\right]
,\ea\ee

Therefore, we have constructed a chiral model on lattice with the knowledge of  
noncommutative differential  
geometry. In the approximation of small parameter $a$, $Tr 
(U_{\mu,\nu}V_{\mu,\nu})\sim TrU_{\mu,\nu}+2 V_{\mu,\nu}$, so most of the 
results are  the same with Smit-Swift model after we freeze 
the radial of higgs field except one term more in the bosonic sector. The naive 
continuum limit of this term is vanish. However, it is needed to be studied 
furture whether this term may lead to any result in physics.  In recent 
years study, it is found that previous chiral models on lattice always meet some 
problems \cite{pet}, so  it is desired to develop new theory to overcome those 
difficulties. We have given a  general rule to construct the chiral model on 
lattice. This may help us to develop the previous models and build new models. 
In this approach, it is also possible to study the chiral models on lattice in 
other schems, such 
as staggered fermion and domail-wall  ect. These topics will be studied  in our 
comming papers.



\centerline{\large \bf Acknowledgements}

This work is supported in part by the National Science Foundation and Chinese 
Post Doctoral Foundation. We  would like to thank Professor H.Y. Guo for many 
useful
discussions. 


\end{document}